\documentclass[preprint,10pt, a4paper]{elsarticle}
\usepackage{amssymb}
\usepackage[draft,cache=True]{minted}
\usepackage{multicol}
\usepackage{lineno}
\usepackage[skins,listings]{tcolorbox}
\usepackage[margin=1in]{geometry}

\journal{SoftwareX}

\usepackage{wrapfig}
\usepackage{color}

\begin{document}

\begin{frontmatter}

\title{A Language and Hardware Independent Approach to Quantum-Classical Computing\tnoteref{t1}}
\tnotetext[t1]{This manuscript has been authored by UT-Battelle, LLC, under contract DE-AC05-00OR22725 with the US Department of Energy (DOE). The US government retains and the publisher, by accepting the article for publication, acknowledges that the US government retains a nonexclusive, paid-up, irrevocable, worldwide license to publish or reproduce the published form of this manuscript, or allow others to do so, for US government purposes. DOE will provide public access to these results of federally sponsored research in accordance with the DOE Public Access Plan. (http://energy.gov/downloads/doe-public-access-plan).}

\author{A. J. McCaskey$^{1,2}$}
\ead{mccaskeyaj@ornl.gov}
\author{E. F. Dumitrescu$^{1,3}$}
\author{D. Liakh$^{1,4}$}
\author{M. Chen$^{5,6}$}
\author{W. Feng$^{6}$}
\author{T. S. Humble$^{1,3,7}$}

\address{$^1$Quantum Computing Institute, Oak Ridge National Laboratory, Oak Ridge, Tennessee, USA}
\address{$^2$Computer Science and Mathematics Division, Oak Ridge National Laboratory,
  Oak Ridge, TN 37831, USA}
\address{$^3$Computational Sciences and Engineering Division, Oak Ridge National Laboratory,
  Oak Ridge, TN 37831, USA}
\address{$^4$Oak Ridge Leadership Computing Facility, Scientific Computing, Oak Ridge National Laboratory, Oak Ridge, Tennessee, USA}
\address{$^5$Department of Physics, Virginia Tech, Blacksburg, VA, USA}
\address{$^6$Department of Computer Science, Virginia Tech, Blacksburg, VA, USA}
\address{$^7$Bredesen Center for Interdisciplinary Research, University of Tennessee, Knoxville, Tennessee, USA}

\begin{abstract}
Heterogeneous high-performance computing (HPC) systems offer novel architectures which accelerate specific workloads through judicious use of specialized coprocessors. A promising architectural approach for future scientific computations is provided by heterogeneous HPC systems integrating quantum processing units (QPUs). To this end, we present \textbf{XACC} (\emph{e\textbf{X}treme-scale \textbf{ACC}elerator}) --- a programming model and software framework that enables quantum acceleration within standard or HPC software workflows. XACC follows a coprocessor machine model that is independent of the underlying quantum computing hardware, thereby enabling quantum programs to be defined and executed on a variety of QPUs types through a unified application programming interface. Moreover, XACC defines a polymorphic low-level intermediate representation, and an extensible compiler frontend that enables language independent quantum programming, thus promoting integration and interoperability across the quantum programming landscape. In this work we define the software architecture enabling our hardware and language independent approach, and demonstrate its usefulness across a range of quantum computing models through illustrative examples involving the compilation and execution of gate and annealing-based quantum programs.
\end{abstract}


\end{frontmatter}


\section{Introduction}
\label{}
High-performance computing (HPC) architectures continue to make strides in the use of specialized computational accelerators, and future HPC designs are expected to increasingly take advantage of compute node heterogeneity \cite{EHReport2018}. Quantum processing units (QPUs) represent a unique coprocessor paradigm which leverages the information-theoretic principles of quantum physics for computational purposes. Several small-scale experimental QPUs, including the publicly available IBM quantum computer \cite{noauthor_ibm_nodate}, already exist and their sophistication, capacity, and reliability continues to improve \cite{Kelly2018}. As a potential HPC accelerator, the emergence of mature QPU technologies requires careful consideration for how to best integrate these devices with conventional computing environments. While the hardware infrastructure for early QPUs is likely to limit their usage to remote access models and state-of-the-art HPC systems \cite{Britt2017}, there are clear use cases where hybrid algorithms may judiciously leverage both conventional and quantum computational resources for near-term scientific applications \cite{deuteron, mcclean_theory_2016}. A hybrid computing paradigm is poised to broadly benefit scientific applications that are ubiquitous within research fields such as modeling and simulation of quantum many-body systems \cite{Kandala2017}, applied numerical mathematics \cite{PhysRevLett.103.150502}, and data analytics \cite{rigettiqaoa}. 
\par
%%%%%%%%%%%%%%%%%%%%%%%%%%%%%
The generalization of HPC programming paradigms to include new accelerators is not without precedent. Integrating graphical processing units (GPUs) into HPC systems was also a challenge for many large-scale scientific applications because of the fundamentally different way programmers interact with the hardware. Hardware-specific solutions provide language extensions \cite{cudaCite} that enable programming natively in the local dialect. Hardware-independent solutions define a hybrid programming specification for offloading work to attached classical accelerators (GPUs, many-integrated core, field-programmable gate array, etc.) in a manner that masks or abstracts the underlying hardware type \cite{openclcite}. These hardware-agnostic approaches have proven useful because they retain a wide degree of flexibility for the programmer by automating those aspects of compilation that are overly complex. Programming models for QPUs will pose additional challenges because of the radically different logical features and physical behaviors of quantum information, such as the no cloning principle and reversible computation. The underlying technology (superconducting, trapped ion, etc.) and models (gate, adiabatic, topological, etc.) will further distinguish QPU accelerators from conventional computing devices. It is therefore necessary to provide flexible classical-quantum programming models and integrating software frameworks to handle the variability of quantum hardware to promote robust application benchmarking and program verification and validation.
\par
%%%%%%%%%%%%%%%%%%%%%%%%%%%%%%
Approaches for interfacing domain computational scientists with quantum computing have progressed over the last few years. A variety of quantum programming languages have been developed with a similar number of efforts under way to implement high-level mechanisms for writing, compiling, and executing quantum code. State-of-the-art approaches provide embedded domain-specific languages for quantum program expression. Examples include the languages and tools from vendors such as Rigetti \cite{quil}, Microsoft \cite{Svore}, Google \cite{cirq}, and IBM \cite{qiskit}, which each enable assembly-level quantum programming alongside existing Pythonic code. Individually, these implementations provide self-contained software stacks that optionally target the vendor's unique hardware implementation or simulator backend. The increasing variability in languages and platforms raises concerns for managing multiple programming environments and compilation tool-chains. The current lack of integration between software stacks increases application development time, decreases portability, and complicates benchmarking analysis. Methods that enable cross-compilation for QPUs will support the broad adoption of experimental quantum computing through faster development time and reusable code. 
\par 
To address these unique challenges, we present a programming model and extensible compiler framework that integrates quantum computing devices into an accelerator-based execution model. The eXtreme-scale ACCelerator (XACC) framework is designed for robust and portable QPU-accelerated application programming by enabling quantum language and hardware interoperability. XACC defines interfaces and abstractions that enable compilation of hybrid programs composed of both conventional and quantum programming languages. The XACC design borrows concepts from existing heterogeneous programming models like OpenCL \cite{openclcite} by providing a hardware-independent interface for off-loading quantum subroutines to a quantum coprocessor. Moreover, XACC enables language interoperability through a low-level quantum intermediate representation. 

The structure of this work is as follows: first, we present related work with regards to quantum programming and detail inherent unique challenges that XACC seeks to address; second, we defined the XACC software architecture, including platform, programming, and memory models; finally, we detail unique demonstrations of the model's flexibility through demonstrations using both gate and annealing quantum computing models.

\section{Related Work}
Programming, compilation, and execution of quantum programs on physical hardware and simulators has progressed rapidly over the last few years. During this time, much research and development has gone into exploring high-level programming languages and compilers \cite{qcl,quipper,qsharp,scaffold}. Moreover, there has been a recent surge in the development of embedded domain specific languages that enable high-level problem expression and automated reduction to low-level quantum assembly languages \cite{qiskit, quil, cirq}. However, despite progress there are still numerous challenges that currently impede adoption of quantum computing within existing classical scientific workflows \cite{McCaskeyICRC2018}. Most approaches that target hardware executions are implemented via Pythonic frameworks that provide data structures for the expression of one and two qubit quantum gates; essentially providing a means for the programming of low-level quantum assembly (QASM). Compiler tools provided as part of these frameworks enable the mapping of an assembly representation to a hardware-specific gate set as well as mapping logical to physical connectivity. The arduous task of complex compiler workflow steps, including efficient instruction scheduling, routing, and robust error mitigation are left as a manual task for the user. This hinders broad adoption of quantum computation by domain computational scientists whose expertise lies outside of quantum information.

Higher-level languages exist, but do not explicitly target any physical hardware. Therefore, users can compile these high-level languages to a representative quantum assembly language, but such instructions must be manually mapped to the set of instructions specified by a given hardware gate set. This translation process is often performed by re-writing the assembly code in terms of a Pythonic execution frameworks targeting a specific device. Moreover, high-level languages have in the past assumed a fault-tolerant perspective of quantum computation. However, this interpretation is at odds with practical near-term noisy computations, for which the user must provide robust compilation tools to enable a variety of error mitigation strategies. To this end, domain specific languages enabling problem expression at higher levels of abstraction \cite{openfermion, acqua, xacc-vqe} for non-fault-tolerant quantum computing have recently been developed. These represent promising pathways for enabling a broad community of computational scientists to benefit from quantum computation.

Overall, currently available quantum languages and compilers are not well integrated with each other. Research and development efforts that offer quantum programming, compilation, and execution mechanisms often target a single simulator or physical hardware instance (i.e. see Pythonic frameworks above). This leads to poor quantum code portability and disables any effort attempting to benchmark various hardware types (superconducting, ion trap, etc.) against each other. Furthermore, there are currently a number of quantum computing models that various research efforts are targeting. A majority of these efforts are targeting the gate model of quantum computation, while others have implemented a noisy form of adiabatic quantum computation. Moreover, there are other models that researchers are becoming increasingly interested in (one-way, topological, etc.). The differences in computational paradigm and hardware specific gate sets negatively affect code portability, verification and validation, and benchmarking efforts.

Our work seeks to address the drawbacks associated with near-term quantum hardware execution and programming models, thus enabling the integration of quantum and classical computation through an extension of the classical coprocessor-computing model. Our aim is to provide a model framework that is extensible to a wide variety of important, practical quantum programming workflow steps. In this way, we can provide a truly integrating quantum compiler and execution framework that works across quantum computing models, languages, and physical (virtual) hardware types. Our efforts aim to benefit quantum code portability, tightly-coupled quantum access models, and classical-quantum application benchmarking efforts.

\section{XACC Architecture}
\label{}
The XACC framework is designed to enable the expression and integration of a wide spectrum of accelerator (quantum) algorithms alongside existing classical code in a manner independent of the accelerator language, hardware, and computational model. Here we define and detail the XACC architecture which we decompose into constituent platform, memory, and programming models. The XACC platform model describes the hardware components at play in the compilation and execution of hybrid programs and how these components behave in relation to one another, while the memory model details the management and movement of data between these components. These model abstractions drive the design and implementation of the XACC programming model, which specifies an application programming interface (API) for offloading computations to an attached quantum accelerator. 

\subsection{Platform and Memory Model}
\label{sec:platform}
XACC treats a general quantum processing unit (QPU) as described in Ref. \citenum{Britt2017}, whereby the QPU is composed of a register of quantum bits (qubits), a quantum control unit (QCU), and a classical memory space for the storage of quantum program results. Ref.~\citenum{Britt2017} puts forth a variety of classical-quantum integration strategies that promote a range of quantum accelerated use cases. For example, one could imagine a loosely-coupled coprocessor model with one or many classical compute nodes accessing a single, remotely hosted QPU. On the other hand, one may consider a tightly-coupled coprocessor model, in which one or many compute nodes have access to an in-memory, in-process device driver API for QPU control and execution (no remote access required). There are, of course, many variants on these models that interpolate between the two extremes. The XACC platform model attempts to take this spectrum into account and enable a variety QPU access models, both local and remote. 

\begin{figure*}[htp]
\includegraphics[width=\textwidth]{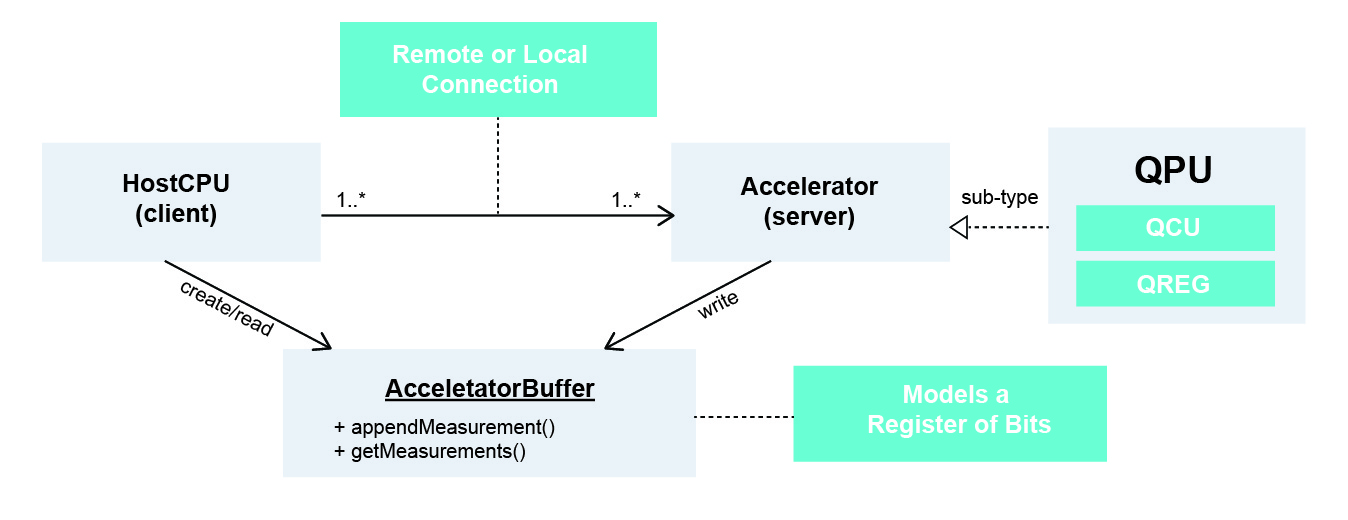}
\centering
\caption{The XACC platform model defines the interplay between the host CPU, accelerator system, and accelerator buffer memory space. The host CPU is charged with judiciously delegating work to a QPU which is controlled by an accelerator system. Results are stored in, and shared by, the accelerator buffer.}
\label{fig:platform}
\end{figure*}

Building on this QPU definition, we define a classical-quantum platform model that enables a range of quantum integration types via the classic \emph{client-server model} \cite{clientserver}, in which programmers of the conventional computing system are on the client side and the quantum accelerator system is on the server side. XACC defines three non-trivial components in this model: (1) the host CPU, (2) the accelerator system (and further sub-types), and (3) the accelerator buffer (see Figure \ref{fig:platform}). The host CPU drives the interaction of classical applications with the attached accelerator system by executing classical applications and delegating quantum computer executions to the attached accelerator system. The role of the accelerator system is to listen for execution requests, and then drive the execution of a quantum program compiled according to the vendor-supplied quantum computer API specifications. The accelerator buffer component forms the underlying hybrid classical-quantum memory space sharing the accelerator execution results.

We have designed XACC to facilitate both serial classical-quantum computing and massively-parallel, distributed high-performance computing enhanced with quantum acceleration, therefore, the cardinality of the host CPU component is one to many (1..*). One could have one or many host CPUs as part of a given hybrid execution corresponding to the many cores available in a HPC application. Likewise, one may consider a computation involving multiple quantum coprocessors. In the very near term this will be unlikely due to QPU infrastructure requirements, but given modest hardware advances the collections of modestly sized QPUs could become available to multiple compute nodes, as is the case with GPUs in classical heterogeneous computing. Therefore, the cardinality of the accelerator system is also one to many (1..*). This platform model allows for the inclusion of multiple classical threads having access to multiple accelerators. 
\par
The XACC memory model ensures that client-side applications can retrieve quantum execution results through the \texttt{AcceleratorBuffer} concept, which models a register of bits and stores ensembles of measurement results. Clients (the host CPU) create instances of the \texttt{AcceleratorBuffer} that are then passed to the accelerator system upon execution. It is the responsibility of the accelerator system to keep track of all measurement results and store them in the \texttt{AcceleratorBuffer}. Since clients keep reference to the created \texttt{AcceleratorBuffer} handle throughout the execution process, the data it contains after execution is available to be post-processed in order to compute expectation values or other statistical quantities of interest, thus influencing the rest of the hybrid computation. 

%%%%%%%%%%%%%%%%%%%%%%%%%%%%%%
\subsection{Programming Model}
The XACC programming model is designed to enable the expression of quantum algorithms alongside existing code in a quantum language-independent manner. Furthermore, the compiled result of the expressed quantum algorithm is designed to be amenable to execution on any quantum hardware through appropriately implemented device drivers. To achieve this, XACC defines six main concepts: (1) accelerator intermediate representation, (2) quantum kernels, (3) transformations on the intermediate representation, (4) compilers, (5) accelerators, and (6) programs. These concepts enable an expressive API for offloading computational tasks to an attached quantum accelerator. Clients express quantum algorithms via quantum kernel source code expressions in a similar way to OpenCL or CUDA for GPUs. These kernels may express quantum algorithms in any quantum programming language for which there exists a valid compiler implementation, thereby enabling a wide variety of programming approaches and techniques (high-level and low-level programmatic abstractions). Compilers map source kernels to a core intermediate representation that enables hardware dependent and independent program analysis, transformations, and optimization. This generally transformed or optimized representation is then mapped to hardware-native assembly code and executed on available physical or virtual hardware instances. 

%In Section 2.1, the authors propose a concept similar to LLVM. In this regard, they state that "To date, there have been no efforts regarding the development of a unified intermediate representation for quantum computing that can span a number of different quantum compute models". However, I would expect a more elaborated discussion about quantum programming language here. The authors are discussing QASM as one example and conclude that this language to too much on the assembly side. But what about the huge amount of other languages which have been proposed for quantum computation in the past (e.g. QCL, LIQUi|>, Quipper, etc.). At the moment, my feeling is that many of them are capable for different levels of abstraction.

\subsubsection{Intermediate Representation}
\label{sec:ir}
%Currently available languages (domain specific and general-purpose) and associated compilers for quantum computing come in a variety of levels of abstraction and provide programmability for a number of different quantum computing models (gate, adiabatic, etc.). 
To promote interoperability and programmability across the wide range of available accelerators and programming languages (embedded or stand-alone), there must exist a low-level program representation that is easy to understand and manipulate. An illustrative example can be found in the LLVM compiler infrastructure which maps various high-level classical programming languages (C, C++, Objective-C, Fortran, etc.) to a common intermediate representation (IR). The IR is then used to perform hardware (dependent and independent) analysis and optimizations in order to generate efficient hardware-specific executable code \cite{llvm}. A standard IR for quantum computation should enable a wide range of programming tools and provide early users the benefit of programming their domain-specific algorithms in a manner that best suits their research and application. To date, there have been no efforts regarding the development of a unified intermediate representation for quantum computing that can span a number of different quantum compute models (e.g., adiabatic, gate). Compiler tools that are currently available take circuit-level programmatic expressions and map them to a hardware-specific quantum assembly (QASM) language, with different efforts providing QASM representations that differ in format and grammar. There is a strong need for a polymorphic set of extendable interfaces that span and support differing quantum accelerator types, thus enabling a retargetable compiler infrastructure for quantum computing across compute models and hardware types. Such an infrastructure sits at a slightly higher level of abstraction than typical assembly representations and therefore enables a unified API that integrates multiple high-level languages with multiple hardware architectures. The goal of this quantum intermediate representation is to provide an assembly-level language and API for quantum program analysis, transformation, and optimization.
%%%%%%%%%%%%%%%%%%%%%%%%%%%%%%
\begin{figure*}
\centering
\includegraphics[width=\textwidth]{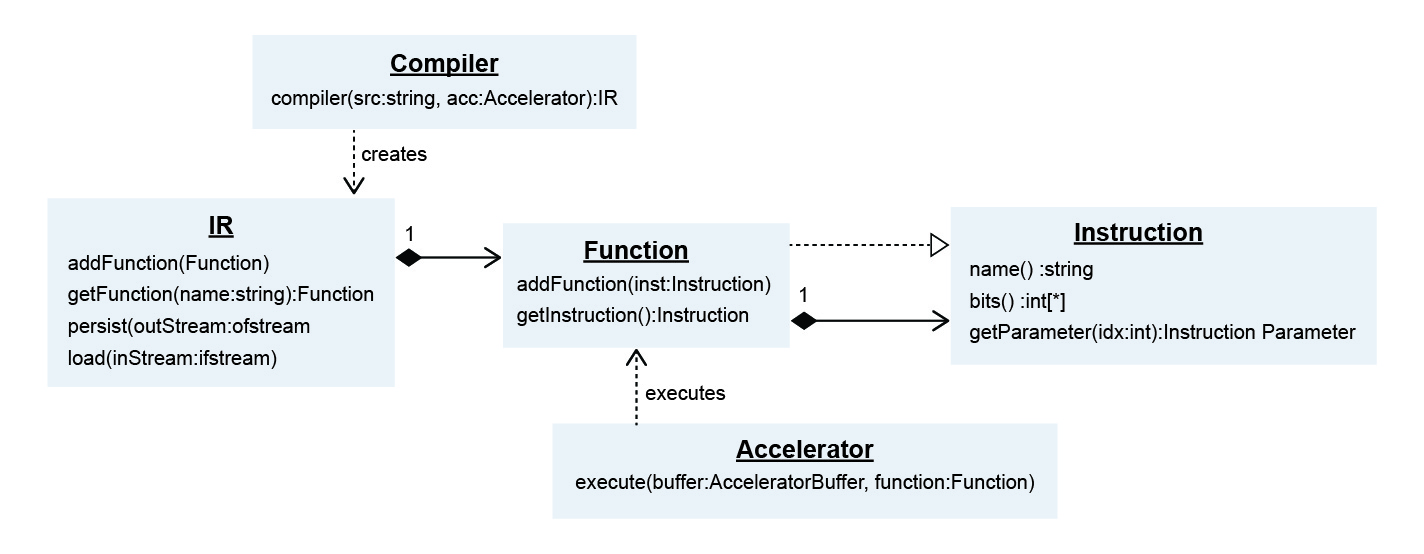}
\caption{The XACC interface architecture. Its core is the \texttt{IR}, \texttt{Function}, and \texttt{Instruction} interfaces and their mutual relationships. \texttt{Compilers} generate compiled \texttt{IR} instances, and \texttt{Accelerators} execute \texttt{IR}-contained \texttt{Functions}. Here solid arrows with a diamond origin imply a composition (\texttt{IR} is composed of \texttt{Functions}), and a dotted line with open-faced arrow implies interface realization (\texttt{Function} \emph{is an} \texttt{Instruction}). }
\label{fig:xacc_arch}
\end{figure*}

\begin{figure*}[htpb]
\centering
\includegraphics[width=\textwidth]{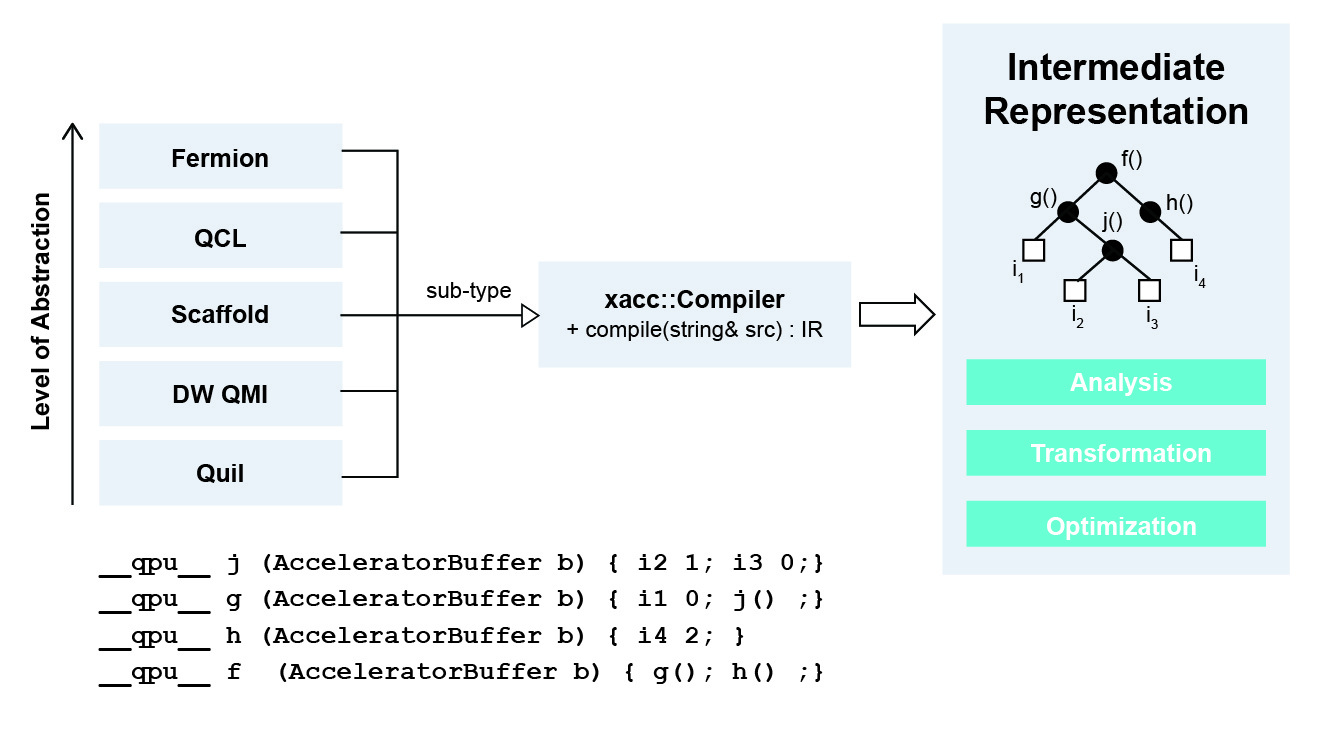}
\caption{The XACC Compiler workflow demonstrating quantum language extensibility. Compilers take quantum kernel source code written in an available language and map it to the XACC intermediate representation for program analysis, transformation, and optimization.}
\label{fig:xacc_flow}
\end{figure*}

XACC defines a polymorphic IR architecture that integrates programming languages and techniques with concrete (physical or virtual) hardware realizations. The XACC IR is designed to adhere to four primary requirements: (1) IR should provide a manipulable in-memory representation and API, (2) IR should be persistable to an on-disk file representation, (3) IR should provide a human-readable, assembly-like representation, and (4) IR should provide a graph representation. The architecture governing the IR interfaces is shown in Figure \ref{fig:xacc_arch} using the Unified Modeling Language (UML). The foundation of the XACC IR is the \texttt{Instruction} interface, which abstracts the concept of an executable instruction (e.g., a quantum gate or program). \texttt{Instructions} have a unique name and reference the accelerator qubits operated upon. \texttt{Instructions} can operate on one or many qubits and can be enabled or disabled for use in classical conditional branching. \texttt{Instructions} can also be parameterized --- each \texttt{Instruction} can optionally keep track of one or many \texttt{InstructionParameters}, which are represented as a variant data structure that can be of type \texttt{float}, \texttt{double}, \texttt{complex}, \texttt{int}, or \texttt{string}. Importantly, the \texttt{InstructionParameter} concept allows a natural representation of instructions in variational quantum algorithms \cite{Peruzzo2014}, which are among some of the most promising candidates for near-term speedups.  
%%%%%%%%%%%%%%%%%%%%%%%%%%%%%%
%%%%%%%%%%%%%%%%%%%%%%%%%%%%%%
Next, XACC defines a \texttt{Function} interface to express source code as compositions of \texttt{Instructions}. The \texttt{Function} interface is a derivation of the \texttt{Instruction} interface that itself contains \texttt{Instructions}. This \texttt{Instruction}/\texttt{Function} combination is an implementation of the composite design pattern, a common software design that models part-whole hierarchies \cite{gof,composite}. Via this pattern, XACC models compiled programs as an \emph{n-ary} tree with \texttt{Function} instances as nodes and \texttt{Instruction} instances as leaves (see Figure \ref{fig:xacc_flow} depicting the mapping between kernel source code and \texttt{Function}/\texttt{Instruction} trees). Executions, transformations, and optimizations of these \texttt{Function} instances are handled via a pre-order tree traversal, whereby walking each node involves walking each child node first, from left to right. For the IR tree in Figure \ref{fig:xacc_flow}, this implies nodes are visited in the following order: $f\rightarrow g\rightarrow i_1\rightarrow j\rightarrow i_2\rightarrow i_3\rightarrow h\rightarrow i_4$, where $f, g, j, h$ are \texttt{Function} instances, and $i_1, i_2, i_3, i_4$ are general \texttt{Instruction} instances. 
Finally, XACC defines the \texttt{IR} interface which serves as a container for \texttt{Functions}. \texttt{IR} contains a list of \texttt{Functions} instances, with an exposed API that enables the mapping of those \texttt{Functions} to both an assembly-like, human-readable string and a graph data structure. For digitized computations, the graph can model the quantum circuit and provides a convenient data structure for program transformation and analysis. For quantum annealing, the graph structure can model the Ising Hamiltonian and scheduling parameters that form the machine-level instructions for the quantum computation. To provide an on-disk representation, the \texttt{IR} interface exposes load and persist methods that take a file path to read in, and to write to, respectively. In this way, \texttt{IR} instances that are generated from a given set of kernels can be persisted and reused, enabling faster ahead-of-time or just-in-time compilation. 

%%%%%%%%%%%%%%%%%%%%%%%%%%%%%%
\subsubsection{IR Transformations}
A key aspect of any compilation workflow is the ability to implement optimizations and transformations, which could be general or hardware dependent. There has been great progress in the development of quantum program transformation, optimization, and optimal instruction scheduling techniques for quantum programs over the last few years \cite{venturelli-temporal, constraint-programming, childs-opt, ibm-mapping}, and we have designed XACC to incorporate such optimizations into its overall compilation workflow. The goal of any program manipulation is to ensure that all compiled instructions are amenable to execution on the desired accelerator in an optimal or near-optimal manner. To handle optimizations and transformations, XACC defines an \texttt{IRTransformation} interface. This interface provides an extension point for taking an \texttt{IR} instance and generating a modified, optimized, or more generally transformed \texttt{IR} instance. The transformed IRs are logically equivalent, i.e., producing equivalent results in an idealized noise-free setting. 

More general IR modifications are particularly well suited to handling error mitigation tasks which are crucial for near-term quantum computations. The basic idea is that on can generate a new IR, or set of transformed IRs, which gather additional information as needed to mitigate against some source of error. In this case, a proper post-execution processing mechanism must be in place to ensure that users retrieve the results they expect. To handle this situation, XACC defines an \texttt{IRPreprocessor} instance to take in the \texttt{IR} instance and modify it in a non-isomorphic manner, but return an \texttt{AcceleratorBufferPostprocessor} instance that knows the details of this modification and can adjust accelerator results accordingly. An example of the utility of this mechanism is in qubit measurement error-mitigation \cite{Kandala2017}, whereby an \texttt{IRPreprocessor} can be implemented that adds measurement kernels to an \texttt{IR} instance. The execution of these additional kernels characterizes readout error rates and can be used by a corresponding \texttt{AcceleratorBufferPostprocessor} implementation, provided by the \texttt{IRPreprocessor} instance, to correct accelerator results. Other mitigation techniques such as noiseless extrapolations and quasi-probability methods \cite{Endo2017, Temme2017} can likewise be handled within the construct of IR pre-procesing and transformations. 

After mapping kernel source code to an IR instance, the XACC model specifies that \texttt{IRTransformations} transform the \texttt{IR} instance before accelerator execution. Following these transformations, all requested or default \texttt{IRPreprocessors} are run and resultant \texttt{AcceleratorBufferPostprocessors} are stored and executed on resultant \texttt{AcceleratorBuffers} after execution.
%%%%%%%%%%%%%%%%%%%%%%%%%%%%%%
\subsubsection{Accelerators}
The inevitable near-term variability in quantum hardware types forces any heterogeneous quantum-classical programming model to be extensible in the hardware it interacts with. XACC is no exception to this and therefore defines an \texttt{Accelerator} interface for injecting physical and virtual (i.e. simulator) QPU backends. The \texttt{Accelerator} interface (shown in Figure \ref{fig:xacc_arch}) provides an \texttt{initialize} operator for sub-types to handle any start-up or loading procedures that are needed before execution on the device. This includes the retrieval of hardware specifications, such as connectivity information, that could influence kernel compilation and \texttt{IR} transformations. \texttt{Accelerators} expose a mechanism for creating \texttt{AcceleratorBuffer} instances, which provide programmers with a handle on \texttt{Accelerator} measurement results. Moreover, \texttt{Accelerator} realizations provide an implementation of a \texttt{getIRTransformations} operation to provide the necessary, low-level hardware-dependent transformations on the logically compiled \texttt{IR} instances. 

Most crucially, \texttt{Accelerators} expose an \texttt{execute} operation that takes as input the \texttt{AcceleratorBuffer} to be operated on and the \texttt{Function} instance representing the kernel to be executed. Realizations of the \texttt{Accelerator} interface are responsible for leveraging these input data structures to affect execution on their target hardware or simulator. It is intended that \texttt{Accelerator} implementations leverage vendor- or library-supplied APIs to perform this execution. All \texttt{execute} implementations are responsible for updating the \texttt{AcceleratorBuffer} with measurement results.  

Note the generality of this \texttt{Accelerator} interface. Subclasses can provide an \texttt{execute} implementation that targets either physical or virtual hardware. In this way we enable available quantum programming languages to target simulated hardware which provides a mechanism for fast feedback on hybrid quantum-classical algorithmic execution. For example, \texttt{Accelerator} developers could provide an \texttt{execute} implementation for a variety of high-performance and specialty simulators \cite{Zulehner2017, Haner2017, qhipster}. In the absence of a preferred simulation methodology, we have provided a default implementation for the \texttt{Accelerator} that enables gate model quantum computer simulation via tensor network theory, specifically through a wave function decomposition leveraging the matrix product state ansatz (Tensor Network Quantum Virtual Machine, TNQVM) \cite{tnqvm-plos-one, tnqvm-github}. This enables users of XACC that target gate model quantum computation to study large systems of qubits before execution on physical hardware (with an upper bound dependent on the level of entanglement in the system). 

\subsubsection{Kernels, Compilers, and Programs}
XACC requires that code intended for an \texttt{Accelerator} be provided in a manner similar to code intended for GPU acceleration within CUDA or OpenCL. That is, code must be expressed via stand-alone \emph{kernels}. A kernel is a programmatic representation of accelerator operations applied to a register of bits. At its core, an XACC kernel is represented by a C-like function, however, this function must take as its first argument the \texttt{AcceleratorBuffer} instance representing the accelerator bit register (qubits) that this kernel operates on. It is in this way that kernels connect classical code with a handle to accelerator measurement results. XACC kernels do not specify a return type; all information about the results of a kernel's operation are gathered from the \texttt{AcceleratorBuffer's} ensemble of bit measurements. Kernels in XACC must be differentiated from conventional library function calls using the \emph{\_\_qpu\_\_} keyword. This annotation can enable static, ahead-of-time compilation of XACC kernels by providing an abstract syntax tree search mechanism. We leave this static, ahead-of-time compiler as future work. Currently, all XACC \texttt{Compilers} are executed at runtime and therefore only enable just-in-time compilation. Finally, Kernels can take any number of kernel arguments that drive the overall execution of the quantum code. These parameters are modeled as the aforementioned \texttt{InstructionParameter} variant type. This enables parameterized compiled \texttt{IR} instances that can be evaluated at runtime. 
\par
\begin{listing}
\begin{minted}
[
frame=lines,
framesep=2mm,
baselinestretch=1.2,
fontsize=\footnotesize,
% linenos
]
{c++}
auto src = R"src(__qpu__ foo(AcceleratorBuffer qreg, double theta) {...})src";
auto qpu = xacc::getAccelerator("ibm");
auto buffer = qpu->createBuffer("qreg", 2);
xacc::Program program(qpu, c
program.build();
auto kernel = program.getKernel<double>("foo");
for (auto& theta : {-3.14...3.14}) kernel(buffer, theta);
\end{minted}
\caption{Example usage of foundational XACC API and the interplay between conventional and quantum programs.}\label{lst:api}
\end{listing}
The function body of an XACC kernel can be expressed in any \emph{available} language. An \emph{available} language is one for which there is a valid \texttt{Compiler} implementation for the language. The \texttt{Compiler} interface architecture is shown in Figure \ref{fig:xacc_arch}, and its extensibility and connection to the XACC IR is shown in Figure \ref{fig:xacc_flow}. This interface provides a \texttt{compile} method that takes kernel source code as input and produces a valid instance of the XACC \texttt{IR}. Derived \texttt{Compilers} are free to perform compilation in any way they see fit, as long as they return a valid \texttt{IR} instance. Moreover, the \texttt{compile} operation can optionally take the targeted accelerator as input, which enables hardware-specific details to be present at compile time and thus influence the way compilation is performed.

This compilation extension point provides a mechanism for the mapping of high-level constructs to lower-level quantum assembly, and therefore facilitates quantum program decomposition methods that map domain specific programmatic expressions to the XACC IR. An example of this would be a domain specific language that expresses a molecular Hamiltonian, and an associated \texttt{Compiler} realization that maps this Hamiltonian to quantum assembly via a Jordan-Wigner or Bravyi-Kitaev transformation \cite{jw,bk}. Note this design also facilitates general gate decomposition techniques through the overall extensibility of the \texttt{compile} method. One could imagine a domain specific language for the expression of general unitaries that are expressed as an XACC kernel and passed to a \texttt{Compiler} implementation that decomposes the unitary into a native low-level gate set \cite{gatedecomp}. 

%Compilers also provide a \texttt{translate} method to enable source-to-source translation. This method takes as input a \texttt{Function} instance and produces an equivalent source string in the \texttt{Compiler's} representative programming language. A typical translation workflow is as follows: a kernel source string can be compiled with its appropriate \texttt{Compiler} instance. The \texttt{Function} instance produced by that operation can then be passed to the \texttt{translate} method of the \texttt{Compiler} for the language being generated. The implementation of the \texttt{translate} method maps the internal \texttt{Instructions} to language-specific source code and returns it. 
\par
The XACC compilation concept also defines a kernel source code \texttt{Preprocessor} extension point. \texttt{Preprocessors} are executed before compilation and take as input the source code to analyze and process, the \texttt{Compiler} reference for the kernel language, and the target accelerator. Using this data, \texttt{Preprocessors} can perform operations on the kernel source string to produce a modified source code that enhances or simplifies a computation. An example of the \texttt{Preprocessor's} utility would be quantum language macro expansion, or searching kernel source code for certain keywords describing a desired algorithm and replacing that line of code with a source-code representation of the algorithm. In this way, \texttt{Preprocessors} can be used to alleviate tedious programming tasks. 
%%%%%%%%%%%%%%%%%%%%%%%%%%%%%%

%%%%%%%%%%%%%%%%%%%%%%%%%%%%%%
%\subsubsection{Programs and Execution Workflow}
The primary entry point for interaction with the XACC compilation infrastructure is the concept of a \texttt{Program}. The \texttt{Program} orchestrates the entire kernel compilation process and provides users with an executable functor to execute the compiled kernel on the desired \texttt{Accelerator}. \texttt{Programs} are instantiated with reference to the kernel source code and targeted \texttt{Accelerator}, and provide programmers with a \texttt{build()} operation that applies requested kernel \texttt{Preprocessors}, selects and executes the correct \texttt{Compiler} to produce the \texttt{IR} instance, and then executes all desired (or default) \texttt{IRTransformations} and \texttt{IRPreprocessors}. Finally, the \texttt{Program} exposes a \texttt{getKernel} operation returning an executable functor that executes the compiled \texttt{Function} on the target \texttt{Accelerator}.

The interplay of conventional and quantum programs is demonstrated in Listing 1. Users describe their source code as an XACC kernel (note this kernel is parameterized by a \texttt{double} parameter), request a reference or handle to the desired \texttt{Accelerator}, and allocate a buffer of qubits. Next, a \texttt{Program} object is instantiated and the XACC compilation workflow is initiated through the \texttt{build} invocation. At this point the appropriate \texttt{Compiler} has mapped the source code to the XACC IR, and all transformations, optimizations, and preprocessors have been invoked to provide an executable functor or lambda that will enable user execution on the desired \texttt{Accelerator}. This executable kernel reference can then be used as part of some parameterized loop, enabling hybrid quantum-classical variational algorithms.

\section{Demonstration}
Near-term quantum computing devices provide a relatively small quantum register and lack sufficient error correction capabilities to implement fault-tolerant computations. Nevertheless, these pre-threshold devices demonstrate sufficient hardware control to support programmable sequences of (imperfect) operations known as quantum circuits. Devices executing these quantum circuits may be used as primitive quantum accelerators within a hybrid computing scheme \cite{AspuruGuzik2015}. Only a few of these early QPUs are publicly available, and all are remotely located with respect to the end user, matching the client-server platform model described in Section \ref{sec:platform}. In this section, we demonstrate the utility of the XACC framework through demonstrations programming both gate and annealing quantum computers, using the unified XACC API. 

\subsection{Example Program for Nuclear Binding Energy Calculations}
\label{}
\par
Here we demonstrate using XACC to compose a scientific application for calculating the binding energy of an atomic nuclei. The accuracy of this program was reported previously for the example of deuteron~\cite{deuteron}, and we use this example to describe the technical details for how this program is constructed. The general structure of this XACC hybrid program derives from the variational quantum eigensolver (VQE) algorithm \cite{mcclean_theory_2016}, which is a quantum-classical algorithm for recovering the lowest energy eigenstate of a quantum mechanical Hamiltonian. The minimal form of the system Hamiltonian, whose lowest eigenvalue is related to the the binding energy, is given by 
\begin{equation}
H_2 = 5.906709 I +0.218291Z_0 -6.125 Z_1 -2.143304 \left(X_0 X_1 + Y_0Y_1\right) ,
\label{eq:deuteron}
\end{equation}
where $X_i, Y_i, Z_i$ denote Pauli operators acting on the $i$th qubit. 
\begin{tcolorbox}[colback=blue!5!white,colframe=blue!75!black,sidebyside, title=Listing 2: XACC Kernels for Deuteron VQE]
\begin{lstlisting}[style=tcblatex]
__qpu__ ansatz(AcceleratorBuffer b, double t0) {
    X 0
    RY(t0) 1
    CNOT 1 0
}
__qpu__ z0(AcceleratorBuffer b, double t0) {
    ansatz(b,t0)
    MEASURE 0 [0]
}
__qpu__ z1(AcceleratorBuffer b, double t0) {
    ansatz(b,t0)
    MEASURE 1 [1]
}
\end{lstlisting}
\tcblower%-----------------------------------------
\begin{lstlisting}[style=tcblatex]
 __qpu__ x0x1(AcceleratorBuffer b, double t0) {
    ansatz(b,t0)
    H 0
    H 1
    MEASURE 0 [0]
    MEASURE 1 [1]
}
__qpu__ y0y1(AcceleratorBuffer b, double t0) {
    ansatz(b,t0)
    RX(1.57079) 0
    RX(1.57079) 1
    MEASURE 0 [0]
    MEASURE 1 [1]
}
\end{lstlisting}
\label{lst:kernels1}
%\caption{Listing 2: XACC Kernels for Deuteron VQE}
\end{tcolorbox}

The VQE algorithm searches for the ground state energy of a given Hamiltonian by optimizing the expectation value of the Hamiltonian with respect to a parameterized quantum wavefunction encoded into the qubit register of an accelerator QPU. For large system sizes, wavefunctions are easily represented in qubit registers but require exponential classical resources to store. At each iteration of this optimization, the QPU is evolved by a quantum circuit parameterized by the current iterate's parameters, and multiple measurements are performed for non-commuting sets of Hamiltonian terms. Expectation values are then evaluated with respect to the ensemble of measurement samples, and the weighted sum of all these expectation values determines the system energy at a given parameterization. This optimization continues until convergence. 

We now demonstrate how to program this algorithm, in a QPU-independent manner, using the XACC framework. First, we define the kernel source code initializing a trial wavefunction, known as an {\em ansatz}, on a QPU. This ansatz is defined subsequently in Listing~2, with the \texttt{\_\_qpu\_\_ ansatz($\cdots$)\{$\cdots$\}} kernel. This kernel implements three logical operations. Using the ansatz kernel as a building block, we append additional  gates and measurement instructions, as needed, to evaluate the expectation values of the Hamiltonian terms from Equation~\ref{eq:deuteron}. The $Z_0$,$Z_1$ terms in Equation \ref{eq:deuteron} can be evaluated with respect to the QPU state after the initialization ansatz, so only measurement instructions are appended. To evaluate the other terms, involving $X$ and $Y$ operators, local change of basis rotations are applied, that is, a Hadamard gate for all $X$ operators and an $X(\frac{\pi}{2})$ rotation for all $Y$ operators. The XACC quantum kernel source code in Listing 2 has been written in Quil \cite{quil}. Note, however, that kernels can be written in any gate model quantum language supported by the framework (OpenQASM, Scaffold, etc.). Because the XACC \texttt{IR} behaves as an n-ary tree of \texttt{Instruction} instances, previously defined kernels can be reused as Instructions in other kernels. Recursive circuits, such as the quantum Fourier transform, can easily be defined in this manner. 
\begin{figure}[!t]
 \begin{minipage}{0.45\textwidth}
  \begin{listing}[H]
  \begin{minted}[
frame=lines,
framesep=2mm,
baselinestretch=1.2,
fontsize=\footnotesize,
% linenos
]{c++}
auto qpu = xacc::getAccelerator("ibm"); 
auto buffer = qpu->createBuffer("qreg", 2);
xacc::Program program(qpu, deuteronSrc);
program.build();
std::vector<double> energies, coeffs{.218291, 
        -6.125, -2.143304, -2.143304};
auto kernels = program.getKernels();
for (auto theta : thetaRange) {
    double energy = 5.906709;
    for (int i = 0; i < kernels.size(); i++) {
        kernels[i](buffer, theta);
        energy += coeffs[i] * 
                buffer->getExpectationValueZ();
        buffer.resetBuffer();
    }
    energies.push_back(energy);
}
\end{minted}
\centering (a)
\end{listing}
\end{minipage}
\begin{minipage}{0.55\textwidth}
\begin{figure}[H]
    % \centering
    \includegraphics[width=\textwidth]{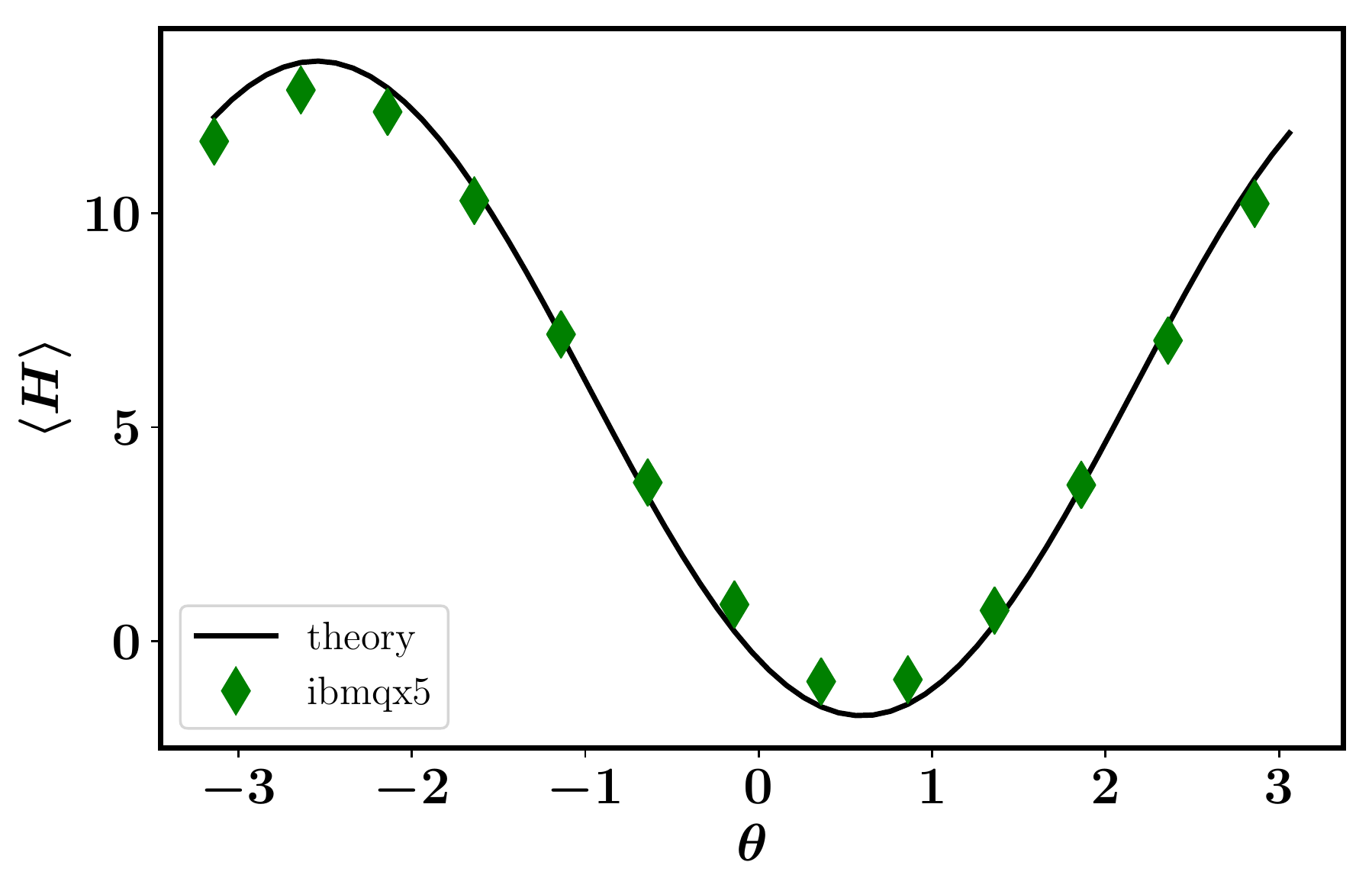}
    \centering (b)
    \label{fig:results}
\end{figure}
 \end{minipage}
 \caption{Parameter variation for deuteron. The code snippet in (a) follows from the code in Listing \ref{lst:api}. Users request an Accelerator and compile XACC kernel source code to XACC IR. Executable kernel functors can be requested and used as part of a parameterized loop. The \texttt{energies} vector as a function of the variational parameter $\theta$ computed via the TNQVM (virtual) Accelerator and IBM QX5 16 qubit QPU  are shown in (b).}
  \label{fig:deuteron}
\end{figure}

To compile and execute these kernels, we leverage the XACC API, as shown in Figure \ref{fig:deuteron} (a). Note that users requesting an \texttt{Accelerator} from the framework simply provide the \texttt{string} name corresponding to the desired \texttt{Accelerator}. This returns a polymorphic \texttt{Accelerator} reference that points to the desired implementation. Running this code on the TNQVM \texttt{Accelerator} amounts to simply modifying the \texttt{getAccelerator} \texttt{string} argument to \texttt{\textbf{tnqvm}}. The results of running this code on the \texttt{TNQVM} \texttt{Accelerator} and the IBMQX5 16 qubit QPU are shown in Figure \ref{fig:deuteron} (b). Raw timing information for this execution is not very illuminating as a large majority of the time is spent waiting in the IBM Quantum Experience job queue or suspect to network lags due to remote HTTPS invocations. We can however estimate a lower bound on the execution times for this example program by considering the circuit length, measurement, and refresh timescales. Let us consider a quantum program consisting of $n_{l(e)}$ layers of local (entangler) quantum gates which may be implemented in parallel (in our example $n_l=2, n_e = 1$). Given typical superconducting gate timescales of $t_l = 20ns, t_e = 200ns$, along with a $t_m=2\mu s$ measurement timescale, and a refresh time of $t_R \approx 10*T_1 \approx 500\mu s$ needed to re-initialize the qubit registers by natural relaxation mechanisms. The minimum device time needed to evaluate all four kernels in Listing~2, given an ensemble size of $10^4$ samples per term, would be 20 seconds for each function evaluation at a given parameter $\theta$. Note that the sample rate can be partially alleviated by parallelization (a topic to be detailed in future work), but one can already see that the overall time resources may become prohibitive (e.g. exponentially costly) for programs which require a significant number of samples in order to optimize over noisy cost function evaluations \cite{Kandala2018, Hempel2018}. It is therefore neccessary to improve the scalability of quantum optimization algorithms in order to reduce the significant cost of quantum optimization. 

%Note that as part of this execution, we have corrected for qubit readout errors via an implementation of the \texttt{IRPreprocessor}. Via a command line option (\texttt{./deuteron --correct-readout-errors}), this implementation gets run after compilation and before execution and acts to append measurement kernels to the \texttt{IR} that enable the computation of readout error probabilities. An \texttt{AcceleratorBufferPostprocessor} implementation then uses these probabilities as part of the post-execution post-processing step to scale and shift computed expectation values. 

\subsection{Simple Integer Prime Factorization on D-Wave}
In an effort to demonstrate the polymorphic nature of the XACC IR, here we provide an example of programming a simple problem targeting the D-Wave QPU. This example leverages the exact same API calls as in Listing 1 and the deuteron demonstration (see Figure \ref{fig:dwfactoring}(b)).
\begin{figure}[!h]
 \begin{minipage}{0.35\textwidth}
  \begin{listing}[H]
  \begin{minted}[
frame=lines,
framesep=2mm,
baselinestretch=1.2,
fontsize=\footnotesize,
% linenos
]{c++}
src = R"src(__qpu__ factor15() {
   0 0 20;
   1 1 50;
   ...
   1 6 -128;
   2 6 -128;
})src";
\end{minted}
\centering (a)
\end{listing}
\end{minipage}
\begin{minipage}{0.55\textwidth}
  \begin{listing}[H]
  \begin{minted}[
frame=lines,
framesep=2mm,
baselinestretch=1.2,
fontsize=\footnotesize,
% linenos
]{c++}
auto qpu = xacc::getAccelerator("dwave");
auto qubitReg = qpu->createBuffer("q");
xacc::Program program(qpu, src);
program.build();
auto factor15 = program.getKernel("factor15");
factor15(qubitReg);
... analyze qubitReg measurement bit strings
\end{minted}
\centering (b)
\end{listing}
 \end{minipage}
 \caption{A simple example demonstrating the flexibility and utility of the XACC framework in executing a program on the D-Wave QPU. (a) represents an XACC kernel written for the D-Wave quantum machine instruction compiler, while (b) demonstrates using the XACC API to compile and execute this code. This code factors 15 into 3 and 5 using the D-Wave QPU.}
  \label{fig:dwfactoring}
\end{figure}
Specifically, we demonstrate the use of an XACC quantum annealing IR implementation (with corresponding \texttt{Function} and \texttt{Instruction} subtypes for quantum annealing) by using the D-Wave QPU to factor 15 into 5 and 3. The quantum kernel for factoring 15 on the D-Wave QPU, and the associated code required to compile and execute it using the XACC API are shown in Figure \ref{fig:dwfactoring} (kernel code trimmed for brevity).

We have implemented a \texttt{Compiler} implementation, the \texttt{DWQMICompiler} \cite{xacc-dwave}, which takes as input kernels structured as a new-line separated list of D-Wave quantum machine instructions (Ising Hamiltonian coefficients). The compilation and execution workflow starts by getting reference to the D-Wave Accelerator, which gives the user access to all remotely hosted D-Wave Solvers (physical and virtual resources). Next, users request that an \texttt{AcceleratorBuffer} be allocated, which gives them a reference to the D-Wave QPU qubits, as well as all resultant data after execution. Then, a \texttt{Program} is created and built (compiled) with reference to the \texttt{Accelerator} and source code. This, in turn begins the minor graph embedding and parameter setting steps as part of the \texttt{DWQMICompiler} workflow (for full details on the D-Wave programming workflow, see \cite{jade}). Users execute the kernel lambda which populates the \texttt{AcceleratorBuffer} instance with the resultant data (energies, measurement bit strings, etc.). The bit string corresponding to the minimum energy can then be used to reconstruct the binary representation of the factors of 15. 

\section{Discussion}
\label{}
We have presented a programming, compilation, and execution framework enabling the integration of quantum computing within standard and HPC workflows in a language and hardware independent manner. We have demonstrated a high-level set of interfaces and programming concepts that support QPU acceleration reminiscent of existing GPU acceleration. These interfaces enable domain computational scientists to migrate existing scientific computing code to early QPU devices while retaining prior programming investments. 

This work opens up interesting avenues for the development of benchmarking, verification, and profiling software suites for near-term quantum computing hardware. As domain computational scientists start leveraging these quantum technologies as part of existing software workflows, the ability to quickly swap out virtual and physical \texttt{Accelerator} instances will enable quick verification of actual QPU results. Benchmarking suites that compare and contrast high-level algorithm executions across the varied quantum hardware types will provide a mechanism for intuiting which hardware best fits the problem at hand. In this regard, XACC provides a unified API for quickly swapping out these hardware instances, thus enabling a write once and run QPU benchmarking and verification mentality. 

Finally, note the generality of the framework's core interfaces. We have focused on the quantum acceleration of classical heterogeneous architectures, but one could easily imagine fitting other post-Moore's law hardware types, such as neuromorphic accelerators, into the XACC framework. This is a direction for future work we intend to pursue.

\section*{Acknowledgements}
\label{}
This work has been supported by the Laboratory Directed Research and Development Program of Oak Ridge National Laboratory, the US Department of Energy (DOE) Office of Science Advanced Scientific Computing Research (ASCR) Early Career Research Award, and the DOE Office of Science ASCR quantum algorithms and testbed programs, under field work proposal numbers ERKJ332 and ERKJ335. This work was also supported by the ORNL Undergraduate Research Participation Program, which is sponsored by ORNL and administered jointly by ORNL and the Oak Ridge Institute for Science and Education (ORISE). ORNL is managed by UT-Battelle, LLC, for the US Department of Energy under contract no. DE-AC05-00OR22725. ORISE is managed by Oak Ridge Associated Universities for the US Department of Energy under contract no. DE-AC05-00OR22750. The US government retains and the publisher, by accepting the article for publication, acknowledges that the US government retains a nonexclusive, paid-up, irrevocable, worldwide license to publish or reproduce the published form of this manuscript, or allow others to do so, for US government purposes. DOE will provide public access to these results of federally sponsored research in accordance with the DOE Public Access Plan.

\section*{References}
%\bibliography{xacc}
\bibliographystyle{elsarticle-num} %iopart-num}

\section*{Required Metadata}
\label{}

\section*{Current code version}
\label{}

\begin{table}[!h]
\begin{tabular}{|l|p{6.5cm}|p{6.5cm}|}
\hline
\textbf{Nr.} & \textbf{Code metadata description} & \textbf{Please fill in this column} \\
\hline
C1 & Current code version & v1.0.0 \\
\hline
C2 & Permanent link to code/repository used for this code version & https://github.com/eclipse/xacc \\
\hline
C3 & Legal Code License   & EPL and EDL\\
\hline
C4 & Code versioning system used & git \\
\hline
C5 & Software code languages, tools, and services used & C++, Python \\
\hline
C6 & Compilation requirements, operating environments \& dependencies & C++11, Boost 1.59+, OpenSSL 1.0.2, CMake\\
\hline
C7 & Link to developer documentation/manual & https://xacc.readthedocs.io \\
\hline
C8 & Support email for questions & xacc-dev@eclipse.org,mccaskeyaj@ornl.gov \\
\hline
\end{tabular}
\caption{Code metadata (mandatory)}
\label{} 
\end{table}

% \section*{Current executable software version}
% \label{}

% Ancillary data table required for sub version of the executable software: (x.1, x.2 etc.) kindly replace examples in right column with the correct information about your executables, and leave the left column as it is.

% \begin{table}[!h]
% \begin{tabular}{|l|p{6.5cm}|p{6.5cm}|}
% \hline
% \textbf{Nr.} & \textbf{(Executable) software metadata description} & \textbf{Please fill in this column} \\
% \hline
% S1 & Current software version & For example 1.1, 2.4 etc. \\
% \hline
% S2 & Permanent link to executables of this version  & For example: $https://github.com/combogenomics/$ $DuctApe/releases/tag/DuctApe-0.16.4$ \\
% \hline
% S3 & Legal Software License & List one of the approved licenses \\
% \hline
% S4 & Computing platforms/Operating Systems & For example Android, BSD, iOS, Linux, OS X, Microsoft Windows, Unix-like , IBM z/OS, distributed/web based etc. \\
% \hline
% S5 & Installation requirements \& dependencies & \\
% \hline
% S6 & If available, link to user manual - if formally published include a reference to the publication in the reference list & For example: $http://mozart.github.io/documentation/$ \\
% \hline
% S7 & Support email for questions & \\
% \hline
% \end{tabular}
% \caption{Software metadata (optional)}
% \label{} 
% \end{table}

\end{document}